\documentclass[aps,floats]{revtex4}
\usepackage{amsmath,amssymb}
\usepackage{graphicx,epsfig}

\begin{document}
\bibliographystyle {plain}

\def\oppropto{\mathop{\propto}} 
\def\opmin{\mathop{\min}} 
\def\opmax{\mathop{\max}} 
\def\opsimeq{\mathop{\simeq}}
\def\opoverderline{\mathop{\overline}}
\def\operarrow{\mathop{\longrightarrow}}
\def\opsim{\mathop{\sim}} 
\def\oplim{\mathop{\lim}}

\def\fig#1#2{\includegraphics[height=#1]{#2}}
\def\figx#1#2{\includegraphics[width=#1]{#2}}


\title{  Random wetting transition on the Cayley tree : \\
a disordered first-order transition  
with two correlation length exponents } 


 \author{ C\'ecile Monthus and Thomas Garel }
  \affiliation{Institut de Physique Th\'{e}orique, CNRS and CEA Saclay
 91191 Gif-sur-Yvette cedex, France}

\begin{abstract}
We consider the random wetting transition on the Cayley tree, i.e. the problem of a directed polymer on the Cayley tree in the presence of random energies along the left-most bonds. In the pure case, there exists a first-order transition between a localized phase and a delocalized phase, with a correlation length exponent $\nu_{pure}=1$. In the disordered case, we find that the transition remains first-order, but that there exists two diverging length scales in the critical region : the typical correlation length diverges with the exponent $\nu_{typ}=1$, whereas the averaged correlation length diverges with the bigger exponent $\nu_{av}=2$ and governs the finite-size scaling properties. We describe the relations with previously studied models that are governed by the same ``Infinite Disorder Fixed Point''. For the present model, where the order parameter is the contact density $\theta_L=l_a/L$ (defined as the ratio of the number $l_a$ of contacts over the total length $L$ ), the notion of ``infinite disorder fixed point'' means that the thermal fluctuations of $\theta_L$ within a given sample, become negligeable at large scale with respect to sample-to-sample fluctuations. We characterize the statistics over the samples of the free-energy and of the contact density. In particular, exactly at criticality, we obtain that the contact density is not self-averaging but remains distributed over the samples in the thermodynamic limit, with the distribution ${\cal P}_{T_c}(\theta) = 1/(\pi \sqrt{ \theta (1-\theta)})$.

\end{abstract}

\maketitle

\section{ Introduction }

In pure phase transitions, the approach to criticality is usually
governed by a single correlation length exponent $\nu$ 
that describes all finite-size scaling properties. 
In the presence of frozen disorder however,
the distribution of correlation functions may become very broad
and disorder-averaged values can become atypical
and dominated by rare events.
This phenomenon has been first studied in one dimensional classical spin
systems \cite{Der_Hil,Cri,luck}, where correlation functions can be
expressed as product of random numbers, and have been  
then found in higher dimensional systems,
such as the 2D random $q$-state Potts model \cite{Lud}
and the random transverse field Ising chain \cite{danielrtfic}
(which is equivalent to the classical 2D McCoy-Wu model).
It is important to stress that both the typical and averaged correlations
 are actually important, depending on the
physical quantities one wants to study \cite{Der_Hil,Cri,danielrtfic}.
It turns out that close to a phase transition, 
the typical correlation length $\xi_{typ}$
and the disorder-averaged correlation length $\xi_{av}$
may have different critical behaviors.
 The best understood example is
 the random transverse field Ising chain,
 which has been studied in great detail by D.S. Fisher 
via a strong disorder renormalization approach
 \cite{danielrtfic} to obtain $ \nu_{typ}=1$ and $\nu_{av}=2$.
These two exponents $ \nu_{typ}=1$ and $\nu_{av}=2$ also occur in 
other disordered models that are described by the same ``Infinite Disorder
Fixed Point'' (see the review \cite{review} and references therein).
In the present paper, we present still another realization of 
this ``Infinite Disorder Fixed Point'' with
$ \nu_{typ}=1$ and $\nu_{av}=2$, as a random wetting transition
on the Cayley tree.

Our physical motivation to consider such a model of random wetting
on the Cayley tree
was to better understand the similarities and differences
with two other types of models involving 
directed polymers and frozen disorder :

(i) for the problem of the directed polymer in a random medium
 on the Cayley tree \cite{Der_Spo}, there exists a freezing transition
towards a low-temperature phase of finite entropy, 
where the polymer is essentially frozen along the random optimal path.
Finite-size properties in the critical region
 \cite{Coo_Der,fssDPCT} have revealed the presence of two distinct
correlation length exponents $\nu=2$ and $\nu'=1$. 
In the random wetting model considered in the present paper,
the difference is that the
the random energies are not on all bonds of the Cayley tree, but only on the
bonds of the left-most path, so that the phase transition corresponds
to a freezing along this boundary path.

(ii) for the wetting \cite{mfisher} and 
Poland-Scheraga model of DNA denaturation \cite{Pol_Scher}
with a loop exponent $c>2$, where the corresponding pure transition
is first-order, we have found numerically \cite{c2.15numerical,PStciL}
that in the presence of frozen
disorder, the transition remains first order, but that 
two correlation length exponents $\nu_{typ}=1$ and $\nu=2$ appear.
The random wetting model considered in the present paper
corresponds to the limit of loop exponent $c \to \infty$ (since loops
do not exist on the Cayley tree) of the model studied in
Ref. \cite{c2.15numerical} 
with the boundary conditions bound-unbound (see more details
on the model in section 2.1 of Ref. \cite{c2.15numerical}).

The paper is organized as follows.
In Section \ref{secmodel}, we introduce 
the model and the interesting observables.
In Section \ref{secfree}, we discuss the statistical properties
of the free-energy and conclude that the wetting transition remains
first-order with a jump of the energy density.
 In section \ref{seclength}, we study the statistics
of the attached and detached lengths, and give the consequences
for the statistics of the order parameter
in the critical region. We summarize our conclusions in \ref{secconclusion}. 
Appendix A contains for comparison the analysis of the properties
of the wetting transition in the pure case. In Appendix B, we discuss
the transition temperatures of the moments of the partition function
in the disordered case.

\section{ Model and observables }

\label{secmodel}

\subsection{ Definition of the model }

\begin{figure}[htbp]
 \includegraphics[height=6cm]{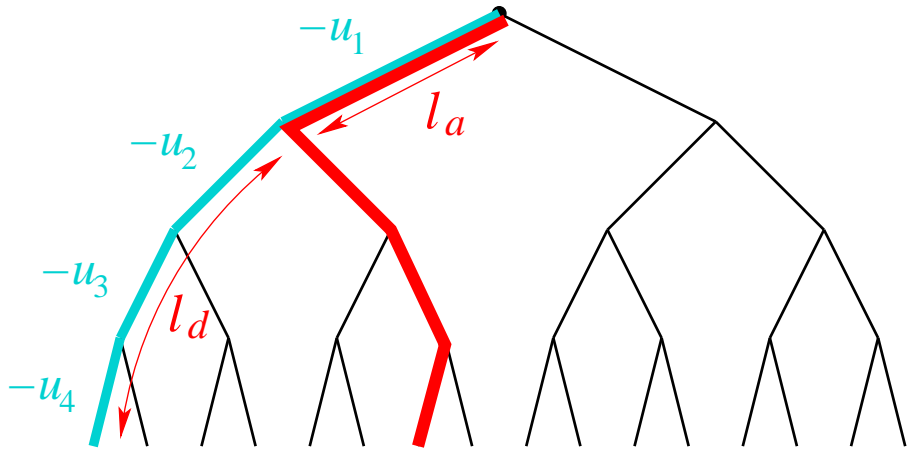}
\caption{ In this paper, we consider a Cayley tree of branching ratio 
$K$ and a number $L$ of generations ($K=2$ and $L=4$ on the Figure),
and we study the statistical physics of a directed polymer starting at
the root in the presence of random energies  $(-u_1,-u_2,..,-u_L)$
along the left-most bonds. A configuration of the directed polymer
is characterized by an attached length $0 \leq l_a \leq L$ along these
left-most bonds and a detached length $l_d=L-l_a$
($l_a=1$ and $l_d=3$ on the Figure). The delocalized phase corresponds to a finite $l_a$ as $L \to +\infty$, whereas the localized phase corresponds to
a finite $l_d$ as $L \to +\infty$. The order parameter is the contact density
$\theta=l_a/L$. }
\label{figtree}
\end{figure}

We consider a directed polymer of length $L$
on a Cayley tree of branching ratio $K$.
The total number of directed walks is simply
\begin{eqnarray}
Y_L=K^L
\label{yL}
\end{eqnarray}
The corresponding delocalized free-energy reads
\begin{eqnarray}
F_L^{deloc}  \equiv  -T \ln Y_L = -T L \ln K
\label{freedeloc}
\end{eqnarray}
and corresponds to an entropy of $(\ln K)$ per bond.

We now consider the random wetting model, 
where the left-most bonds are characterized by 
independent random energies $(-u_i)$
with $i=1,2,..,L$ (see Figure \ref{figtree}).
As an example, they can be drawn from the Gaussian distribution
of average value $u_0>0$ and variance $\Delta^2$ 
\begin{eqnarray}
p(u) = \frac{1}{\sqrt{2 \pi \Delta^2}} e^{ -\frac{(u_i-u_0)^2}{2 \Delta^2}}
\label{gauss}
\end{eqnarray}

\subsection{ Recurrence satisfied by the partition function  }

To write a recurrence for the partition function,
it is convenient to label the random energies from the bottom,
instead of the labelling from the root shown on Fig. \ref{figtree}
\begin{eqnarray}
{\tilde u}_i = u_{L-i}
\end{eqnarray}

The partition function $Z_L$ then
satisfies the simple recurrence 
\begin{eqnarray}
Z_{L+1}= e^{\beta {\tilde u}_{L+1} } Z_L +(K-1) Y_L
\label{reczl}
\end{eqnarray}
with the initial condition $Z_0=1$.
It is convenient to introduce the ratio 
with respect to the free partition function of Eq. \ref{yL}
\begin{eqnarray}
R_L \equiv \frac{Z_L}{Y_L} = \frac{Z_L}{K^L}
\label{defRL}
\end{eqnarray}
The recurrence then becomes 
\begin{eqnarray}
R_{L+1}= \frac{e^{\beta {\tilde u}_{L+1} }}{K} R_L +\frac{(K-1) }{K}
\label{recRL}
\end{eqnarray}
with the initial condition $R_0=1$.

\subsection{ Free-energy difference with the delocalized state }

The difference between the free-energy $F_L=-T \ln Z_L$ 
and the delocalized free-energy of Eq. \ref{freedeloc}
is given by
\begin{eqnarray}
F_L^{diff} \equiv F_L- F_L^{deloc}  = -T \ln Z_L +T \ln Y_L = -T  \ln R_L
\label{freediff}
\end{eqnarray}
The excess free-energy per monomer due to the wall
 in the thermodynamic limit $L \to +\infty$
\begin{eqnarray}
f_{\infty} \equiv \oplim_{L \to + \infty} \frac{ F_L- F_L^{deloc} }{L} =  -T  \oplim_{L \to + \infty} \frac{ \ln R_L}{L} 
\label{freediffintensif}
\end{eqnarray}
characterizes the wetting transition : 
it vanishes in the delocalized phase $f_{\infty}(T>T_c) =0$
and remains finite in the localized phase $f_{\infty}(T<T_c) <0$.

\subsection{ Probability distributions of the attached length $l_a$
and of the detached length $l_d$  }

Since there is no loop on the Cayley tree, 
the directed polymer cannot return to the wall
after leaving it. 
A configuration can be thus decomposed into 
a length $l_a$ attached to the wall starting at the root
and a length $l_d=L-l_a$ detached from the wall (see Fig. \ref{figtree}).
The thermal probability $Q_L(l_a)$
to have exactly $l_a$ links attached to the wall in a given disordered sample
reads
\begin{eqnarray}
Q_L(l_a)   = \frac{ 
e^{ \displaystyle \beta \sum_{i=1}^{l_a} {u}_i}  \left[ \delta_{la,L} 
+ (K-1) K^{L-l_a-1} \theta(l_a <L) \right] }{ Z_L} 
\label{Qla}
\end{eqnarray}
 The numerator is the contribution
 to the partition function of the configurations
that leave the wall after exactly $l_a$ steps.
The denominator is the partition function
\begin{eqnarray}
Z_L =  \sum_{l_a=0}^L
e^{ \displaystyle \beta \sum_{i=1}^{l_a} { u}_i}  \left[ \delta_{la,L} 
+ (K-1) K^{L-l_a-1} \theta(l_a <L) \right]  
\label{ZLQla}
\end{eqnarray}
so that the normalization of the probability distribution $Q_L(l_a)$ reads 
\begin{eqnarray}
\sum_{l_a=0}^L Q_L(l_a) = 1
\label{normapdetached}
\end{eqnarray}
It will be convenient to introduce also 
the probability distribution of the detached length $l_d=L-l_a$
\begin{eqnarray}
P_L(l_d)  \equiv Q_L(l_a=L-l_d ) = \frac{ 
e^{  \displaystyle \beta \sum_{i=1}^{L-l_d} u_i} 
 \left[ \delta_{l_d,0} + (K-1) K^{l_d-1} \theta(l_d \geq 1) \right] 
}{ Z_L} 
\label{pdetached}
\end{eqnarray}

\subsection{ Order parameter : contact density}

A convenient order parameter of the transition is the contact density,
which is directly related to the attached length $l_a$
 introduced in Eq. \ref{Qla}
\begin{eqnarray}
\theta_L \equiv \frac{l_a}{L}
\label{thetal}
\end{eqnarray}
In particular, its thermal average in a given disordered sample 
is determined by the first moment of the distribution $Q_L(l_a)$
of Eq. \ref{Qla}
\begin{eqnarray}
< \theta_L > = \frac{ <l_a > }{L} =
 \frac{1}{L} \sum_{l_a=0}^{L} l_a Q_L(l_a) 
 \label{thermalavthetal}
\end{eqnarray}
In the thermodynamic limit $L \to \infty$, it is expected to remain finite 
in the localized phase and to vanish
in the delocalized phase.

\subsection{Internal Energy  } 

The internal energy 
\begin{eqnarray}
E_L = -\partial_{\beta} \ln Z_L = -\partial_{\beta} \ln R_L
\label{enerdef}
\end{eqnarray}
satisfies the recursion (see Eq. \ref{recRL} for the corresponding recursion
for the reduced partition function $R_L$)
\begin{eqnarray}
 E_{L+1}  = \frac{e^{\beta {\tilde u}_{L+1} } R_L (- {\tilde u}_{L+1}+E_L)}
{e^{\beta {\tilde u}_{L+1} } R_L +(K-1) }
\label{recener}
\end{eqnarray}
and the initial condition $E_0=0$.
In terms of the attached length $l_a$ introduced above 
 (see Eq. \ref{Qla}),
it can also be written as 
\begin{eqnarray}
 E_L  = \frac{ \displaystyle \sum_{l_a=0}^{L} \left( \displaystyle \sum_{i=1}^{l_a} u_i \right)
e^{ \displaystyle \beta \sum_{i=1}^{l_a} u_i}  \left[ \delta_{la,L} 
+ (K-1) K^{L-l_a-1} \theta(l_a <L) \right] 
}{ Z_L} 
\label{enerla}
\end{eqnarray}

The energy per monomer in the thermodynamic limit $L \to +\infty$
\begin{eqnarray}
e_{\infty} \equiv \oplim_{L \to + \infty} \frac{ E_L }{L}
\label{esintensif}
\end{eqnarray}
remains finite in the localized phase $e_{\infty} (T<T_c)>0$
and vanishes in the delocalized phase $e_{\infty} (T>T_c)=0$.
Its value can be obtained via the usual thermodynamic relation from 
the excess free-energy $f_{\infty}$ introduced in Eq. \ref{freediffintensif} 
\begin{eqnarray}
e_{\infty} = \partial_{\beta} ( \beta f_{\infty} ) 
\label{thermoef}
\end{eqnarray}

\section{ Statistical properties of the free-energy   }

\label{secfree}

In this section, we discuss the statistics of the free-energy difference 
of Eq. \ref{freediff} which depends only on the variable
$R_L$ introduced in \ref{defRL}. We first explain
that $R_L$ is a Kesten random variable and 
describe its statistical properties.

\subsection{ Analysis of the variable $R_L$ as a Kesten random variable  }

The recursion of Eq. \ref{recRL} takes the form
 of the recurrence of Kesten variable  \cite{kes75} 
\begin{eqnarray}
R_{L+1}=  a_{L+1} R_L +b
\label{reckesten}
\end{eqnarray}
where $b=(K-1)/K$ is a constant and where the
\begin{eqnarray}
a_{i}= \frac{e^{\beta u_{i} }}{K} 
\label{aLkesten}
\end{eqnarray}
are independent identically distributed random variables.
The specific structure of a Kesten variable $R_L$ 
consists in a sum of products of random variables
\begin{eqnarray}
R_{L}=  \prod_{i=1}^L a_i + b \left( 1+ \sum_{j=2}^{L} \prod_{i=j}^L a_i \right)
\label{discreteKesten}
\end{eqnarray}
This discrete form has for continuous analog the exponential functional 
\begin{eqnarray}
{\cal R}_L= \int_0^L dx e^{ \int_x^L dy F(y) }
\label{kestencontinuous}
\end{eqnarray}
where $\{F(x)\}$ is the random process 
corresponding to the random variables $(\ln a_i)$
in the continuous limit.

This type of random variables, either in the discrete or continuous forms,
 appears in a variety of disordered systems,
in particular in random walks in random media
 \cite{sol75,sinai,derrida,afa90,bou90,
bur92,oshanin,flux,us_sinai,us_golosov},
in the classical random field Ising chain \cite{hilhorst,calan},
and in the quantum random transverse field Ising chain 
\cite{igloi98,dharyoung,microcano}.
Many properties have been thus already studied in great detail
in these previous works. In the following, we cite and translate
these known results for the present context.

To describe the universal results near the critical point
for large samples, there are only two relevant parameters :

(i) the first important parameter is the averaged value 
\begin{eqnarray}
F_0 \equiv \overline{ \ln a_i } = \beta \overline{u}  - \ln K 
\label{defF0}
\end{eqnarray}
 The critical point corresponds to the vanishing condition
$ \overline{ \ln a_i }=0$, so that the critical temperature 
of the present random wetting model reads
\begin{eqnarray}
T_c = \frac{ \overline{u}  }{ \ln K }
\label{tcdesordre}
\end{eqnarray}
(In particular, for the Gaussian distribution of Eq. \ref{gauss}, 
the critical temperature is
 $T_c=u_0/(\ln K)$ and is independent of the variance $\Delta^2$.)

(ii) the second important parameter is the variance of $(\ln a_i)$
that will be denoted by 
\begin{eqnarray}
2 \sigma \equiv \overline{ (\ln a_i)^2 } - (\overline{ \ln a_i })^2 = 
\beta \left( \overline{ u_i^2 } - (\overline{u_i})^2\right)
\label{defsigma}
\end{eqnarray}
to keep the same notations as in Refs. \cite{flux,us_golosov,microcano}.
In particular, from the Central Theorem,
the sum $\Sigma_N = \displaystyle \sum_{i=1}^N (\ln a_i)$ 
is then distributed asymptotically for large $N$
with the Gaussian distribution
\begin{eqnarray}
P_N(\Sigma_N) \opsimeq_{N \to \infty} \frac{1}{\sqrt{4 \pi \sigma N}} e^{ - \frac{(\Sigma_N- N F_0)^2}{4 \sigma N}}
\label{CLT}
\end{eqnarray}

\subsection{ Delocalized phase $T>T_c$  }

For $T>T_c$, the parameter of Eq. \ref{defF0}
is negative $F_0<0$ 
and the random variable $R_L$ remains a finite random
 variable as $L \to \infty$.
Its probability distribution $P_{\infty} (R_{\infty} )$
 is known to display the power-law decay
\cite{kes75,hilhorst,calan} 
\begin{eqnarray}
P_{\infty} (R_{\infty} ) \oppropto_{ R_{\infty} \to \infty } \frac{1}{R_{\infty}^{1+\mu(T)}  }
\label{powermu}
\end{eqnarray}
where the exponent $\mu(T)>0$ is determined by the condition
\begin{eqnarray}
1 = \overline{ a^{\mu}_i } 
= \frac{ \overline{ e^{\mu \beta u_i }}}{K^{\mu}} 
\label{eqmu}
\end{eqnarray}

In conclusion, the difference of Eq. \ref{freediff}
between the free-energy $F_L$ and the delocalized free-energy
$F_L^{deloc}$ remains a finite random variable as $L \to \infty$
\begin{eqnarray}
F_L^{diff} \equiv F_L - F_L^{deloc}
 \opsimeq_{L \to \infty} F^{diff} \equiv -T \ln R_{\infty}
\label{fdiffdeloc}
\end{eqnarray}
and its distribution presents the following
exponential tail for $F_{diff} \to -\infty$
(see Eq. \ref{powermu}) 
\begin{eqnarray}
P_{deloc}(F^{diff}) \opsimeq_{F_{diff} \to - \infty} e^{ \beta \mu(T) F_{diff} }
\label{pfdiffdeloc}
\end{eqnarray}

Example : for the case of the Gaussian distribution of 
Eq. \ref{gauss}, the condition of Eq. \ref{eqmu}
 becomes
 \begin{eqnarray}
1 = \frac{  e^{\mu \beta u_{0} 
 + \frac{\mu^2 \beta^2 \Delta^2}{2} }}{K^{\mu}} 
\label{eqmugauss}
\end{eqnarray}
 and the exponent $\mu$ reads
\begin{eqnarray}
\mu^{Gauss} (T) = \frac{2 T (T \ln K - \overline{u_{L}}) }{  \Delta^2 }
 = \frac{2 (\ln K) T (T - T_c) }{  \Delta^2 } 
\label{resmugauss}
\end{eqnarray}

\subsection{ Critical point $T=T_c$  }

Exactly at criticality $T=T_c$, 
the parameter of Eq. \ref{defF0} vanishes $F_0=0$.
It is then known (see for instance \cite{flux,yor,microcano} 
and references therein) that the leading behavior is of order
\begin{eqnarray}
\ln R_L(T_c) \opsimeq_{L \to \infty} 2 (\sigma L)^{1/2} w_c
\label{freecriti}
\end{eqnarray}
where $w_c$ is a positive random variable of order $O(1)$.
Moreover, it can be shown (see \cite{microcano} 
and references therein) that the distribution of $w_c$ is 
the half-Gaussian distribution 
(cf Eqs (24) and (32) in \cite{microcano})
\begin{eqnarray}
P(w_c) \simeq \theta(w_c \geq 0) \frac{2}{\sqrt \pi} e^{- w^2}
\label{pwtc}
\end{eqnarray}

\subsection{ Localized phase $T<T_c$  }

For $T<T_c$, the parameter of Eq. \ref{defF0} is positive $F_0>0$ 
and the random variable $R_L$ will grow exponentially  in $L$
\begin{eqnarray}
\ln R_L(T<T_c) \opsimeq_{L \to \infty} L F_0 
+   2 \sqrt {\sigma L} w
\label{freeloc}
\end{eqnarray}
The random variable $w$ of order $O(1)$ is distributed with the 
Gaussian distribution 
(cf Eqs (24) and (31) in \cite{microcano})
\begin{eqnarray}
P(w) \simeq  \frac{1}{\sqrt \pi} e^{- w^2}
\label{pwloc}
\end{eqnarray}

\subsection{ Summary of the critical behavior of the free-energy
and energy  }

In conclusion, the difference of Eq. \ref{freediff}
between the free-energy $F_L$ and the delocalized free-energy
$F_L^{deloc}$ displays the following critical behavior

\begin{eqnarray}
 F_L^{diff} (T>T_c) && =
 F_{diff} \ \ {\rm  of \ order} \  O(1) \nonumber \\
 F_L^{diff} (T_c)  && =- T_c 2 (\sigma L)^{1/2} w_c
 \ \ \ \ {\rm with } \  w_c>0 \ {\rm   of \ order} \  O(1) \nonumber \\
 F_L^{diff} (T<T_c) && =
- L ( \overline{u}  - T \ln K ) 
 -T    2 \sqrt {\sigma L} w
 \ \ \ \ {\rm with } \  w \ {\rm  of \ order} \  O(1)
\label{bilanfree}
\end{eqnarray}
where the distributions of the random variables $F_{diff}$, $w_c$ and $w$
have been given above in Eqs \ref{pfdiffdeloc}, \ref{pwtc} and \ref{pwloc}.
In the thermodynamic limit $L \to +\infty$,
the excess free-energy per monomer due to the wall
 defined in Eq. \ref{freediffintensif} is thus given by
\begin{eqnarray}
f_{\infty} (T \geq T_c) && =0 \\
f_{\infty} (T \leq T_c) && = -  ( \overline{u}  - T \ln K ) = - (T_c-T) \ln K
\label{resfreediffintensif}
\end{eqnarray}
The thermodynamic relation of Eq. \ref{thermoef} then gives the behavior
of the energy per monomer
\begin{eqnarray}
e_{\infty} (T > T_c) && =0 \\
e_{\infty} (T < T_c) && =  - T_c \ln K  = -  \overline{u} 
\label{resenerintensif}
\end{eqnarray}
The transition is thus first-order, with a jump of the energy per monomer.

\section{ Statistical properties of the attached and detached lengths  }

\label{seclength}

In this section, we discuss 
the statistical properties of the attached length $l_a$
or of the complementary detached length $l_d$ whose distributions have been introduced in Eq. \ref{Qla} and Eq. \ref{pdetached} respectively.
This allows to analyse also the statistics of the 
 contact density of Eq. \ref{thetal}, which represents the order 
parameter and of the internal energy of Eq. \ref{enerla}. 

\subsection{ Typical behavior of $Q_L(l_a)$ in the delocalized phase }

In the delocalized phase, the attached length $l_a$
will remain a finite random variable as $L \to \infty$.
The typical decay of $Q_L(l_a)$ is then given by (Eq. \ref{Qla})
\begin{eqnarray}
\overline{ \ln Q_L(l_a) } && =  \beta l_a \overline{ u} 
+ (L-l_a) \ln K + \ln \frac{K-1}{K} - \overline{ \ln  Z_L} \\
&& \simeq - l_a (  \ln K - \beta \overline{ u} ) 
+ \ln \frac{K-1}{K} + (L \ln K - \overline{ \ln  Z_L})
\label{qatyp}
\end{eqnarray}
i.e. the decay is governed by the typical correlation length 
\begin{eqnarray}
\frac{1}{\xi_{typ} } =  \ln K - \beta \overline{ u}  
\label{xitypqa}
\end{eqnarray}
that diverges with the correlation length exponent 
\begin{eqnarray}
\nu_{typ} =1
\label{nutydeloc}
\end{eqnarray}
as in the pure case ( $\nu_{pur}=1$ of Eq. \ref{xipuredeloc}).

\subsection{ Typical decay of $P_L(l_d)$ in the localized phase }

In the localized phase, the detached length $l_d$ will remain a finite
random variable as $L \to +\infty$. The typical decay of $P_L(l_d)$
is then given by
\begin{eqnarray}
\overline{ \ln P_L(l_d) } && =  \beta (L-l_d) \overline{ u} 
+ l_d \ln K + \ln \frac{K-1}{K} - \overline{ \ln  Z_L} \\
&& \simeq - l_d ( \beta \overline{ u} - \ln K ) + \ln \frac{K-1}{K}
+ (L \beta \overline{ u} - \overline{ \ln  Z_L} )
\label{pdetachedtyp}
\end{eqnarray}
i.e. the decay is governed by the typical correlation length
\begin{eqnarray}
\frac{1}{\xi_{typ} } =  \beta \overline{ u} - \ln K 
\label{xityppopon}
\end{eqnarray}
that diverges with the correlation length exponent 
\begin{eqnarray}
\nu_{typ} =1
\label{nutyloc}
\end{eqnarray}
as in the pure case ( $\nu_{pur}=1$ of Eq. \ref{xipureloc}).

\subsection{ Statistics over the samples
of the probability $q_L=Q_L(l_a=0)$ of zero contacts }

To understand why typical and disordered averaged behaviors
can be different for the distributions $Q_L(l_a)$ and $P_L(l_d)$,
it is convenient to consider first the particular case
of the probability of zero contacts
\begin{eqnarray}
q_L \equiv Q_L(l_a=0)= P_L(l_d=L) = 
\frac{ (K-1) K^{L-1} }{ Z_L} = \frac{(K-1)}{K } \frac{1}{R_L} 
\label{qLzero}
\end{eqnarray}
because it depends only on the Kesten variable $R_L$
whose statistics has been discussed in detail in section \ref{secfree}.
We also note that $q_L$ coincides with the expression of the flux $J_L$
in the Sinai model, and we refer to \cite{flux} 
where many exact results have been derived for its probability distribution 
(see in particular Eq. (6.10) in \cite{flux}).
In the following, we concentrate on the critical region.
From the statistics of $R_L$ described in section \ref{secfree},
one has in particular that

(i)  in the delocalized phase, $q_L$ remains a finite random variable
as $L \to +\infty$

(ii) in the localized phase, the probability $q_L$ becomes exponentially small
in $L$ 

(ii) exactly at criticality, the variable $q_L$ behaves as
$q_L \sim e^{- w_c 2 \sqrt {\sigma L}}$ where 
$w_c$ is a positive random variable of order $O(1)$ distributed with the 
half-Gaussian distribution of Eq. \ref{pwtc}.
As discussed in \cite{microcano}, an important consequence of the
critical statistics of Eq.\ref{pwtc}
is that the averaged value is dominated by the rare samples having
an anomalously small value of the scaling variable $0 \leq w \leq 1/\sqrt L$,
and since the probability distribution $P(w_c)$ is finite at $w_c=0$,
one obtains the slow power-law decay
\begin{eqnarray}
\overline{ q_L(T_c) } \oppropto_{L \to \infty} \frac{1}{\sqrt L}
\label{rarepwtc}
\end{eqnarray}
whereas the typical behavior is given by the exponential decay
\begin{eqnarray}
\overline{ \ln q_L(T_c) } \oppropto_{L \to \infty} - \sqrt L
\label{typpwtc}
\end{eqnarray}

We refer to \cite{microcano} and references therein for
the discussion of similar examples of averaged values dominated by rare events
in the field of quantum spin chains.

\subsection{ Notion of `` Infinite Disorder Fixed Point''  }

The probability distribution $Q_L(l_a)$ of the attached length $l_a$
in a given sample has the form
of a Boltzmann measure 
\begin{eqnarray}
Q_L(l_a) =  \frac{  e^{- U(l_a)}  }{ \displaystyle
\sum_{l_a=0}^L e^{- U(l_a)}}
\label{Boltzmann}
\end{eqnarray}
over the random walk potential 
\begin{eqnarray}
U(l_a) =  - \beta \sum_{i=1}^{l_a} { u}_i + l_a \ln K= - \sum_{i=1}^{l_a} \ln a_i
\label{Uladiscrete}
\end{eqnarray}
The continuum analog of Eq. \ref{Boltzmann} is then the Boltzmann measure
over a Brownian potential (see Eq. 10 in  \cite{us_golosov}).
We refer to \cite{us_golosov} and references therein 
for more detailed discussions, but the essential property
 is that the Boltzmann measure over a Brownian valley is very concentrated
(in a region of order $O(1)$) around the absolute minimum
of the Brownian potential. Within our present notations, this means 
that in a given disordered sample at a given temperature,
the thermal probability distribution $Q_L(l_a)$
is concentrated in a finite region
around the point $l_a^*$ where the random potential
of Eq. \ref{Uladiscrete}
 reaches its minimum on the interval $0 \leq l_a \leq L$
\begin{eqnarray}
Q_L(l_a) \simeq \delta(l_a-l_a^*)
\label{concentration}
\end{eqnarray}
This means that in a given sample, the contact density of Eq. \ref{thetal}
is essentially given by its thermal average
\begin{eqnarray}
\theta_L \simeq < \theta_L > = \frac{l_a^*}{L}
\label{thetaletav}
\end{eqnarray}

\subsection{ Sample-to-sample fluctuations of the contact density }

We now discuss the sample-to-sample fluctuations of $l_a^*$ and of the
contact density.
The disorder average of the probability distribution $Q_L(l_a)$
can be obtained as
\begin{eqnarray}
\overline{ Q_L(l_a) } \simeq \overline{ \delta (l_a-l_a^*) } = \pi_L(l_a)
\label{avQLa}
\end{eqnarray}
where $\pi_L(l_a^*)$ is the probability distribution 
of the minimum $l_a^*$ over the samples.

Taking into account the two parameters $F_0$ (Eq. \ref{defF0}) and $\sigma$
(Eq. \ref{defsigma}) that characterize the large scale properties of the 
random walk $U(l_a)$, 
the probability distribution $\pi_L(l_a^*)$ of the minimum $l_a^*$
can be obtained as 
\begin{eqnarray}
\pi_L(l_a^*) = \frac{\psi^{(F_0)}(L-l_a^*)  \psi^{(-F_0)} (l_a^*) }
{ \int_0^L dl \psi^{(F_0)}(L-l_a^*)  \psi^{(-F_0)} (l_a^*)   }
\label{piG}
\end{eqnarray}
where 
\begin{eqnarray}
\psi^{(F_0)} (l) = \int_0^{+\infty} dU G^{(F_0)} (U,l)
\label{defpsi}
\end{eqnarray}
and where $G^{(F_0)} (U, l)$ represents the probability for a biased 
random walk to go
from $(0,0)$ to $(U,l)$ in the presence of an absorbing wall at $U=0^-$
(see for instance \cite{microcano} for very similar calculations)
\begin{eqnarray}
G^{(F_0)} (U, l) = \frac{ U }{ 2 {\sqrt \pi} (\sigma l)^{3/2}}
e^{ - \frac{ (U- F_0 l)^2}{4 \sigma l} }
\label{defG}
\end{eqnarray}
The critical behavior can be analyzed as follows.

\subsubsection{ Sample-to-sample fluctuations at criticality }

 At criticality where $F_0=0$, the function $\psi$ of Eq. \ref{defpsi}
reads
\begin{eqnarray}
\psi^{F_0=0} (l) = \int_0^{+\infty} dU  \frac{ U }
{ 2 {\sqrt \pi} (\sigma l)^{3/2} }
e^{ - \frac{ U^2}{4 \sigma l} } = \frac{1}{ \sqrt{ \pi \sigma l} }
\label{psicriti}
\end{eqnarray}
The distribution of Eq. \ref{piG} for the position $l_a^*$ then reads
\begin{eqnarray}
\overline{ Q_L(l_a^*) } = \pi_L^{(T_c)}(l_a^*) = \frac{1}{ \pi \sqrt{ l_a^* (L-l_a^*)} }
\label{pilatc}
\end{eqnarray}

As a consequence, we find that the 
contact density $\theta  \simeq l_a^*/L$
is distributed at criticality over the samples with the law
\begin{eqnarray}
{\cal P}_{T_c}(\theta) = 
\frac{1}{ \pi \sqrt{ \theta (1- \theta)} }
\label{contacttc}
\end{eqnarray}

\subsubsection{ Sample-to-sample fluctuations in the delocalized phase }

In the delocalized phase where $F_0<0$,
the position $l_a^*$ of the minimum on $[0,L]$
of the Brownian potential of positive drift remains finite as $L \to +\infty$,
and the probability distribution of Eq. \ref{piG} can be obtained as
\begin{eqnarray}
\pi_L^{(T>T_c)}(l_a^*) \simeq_{L \to \infty}   \frac{\psi^{(-F_0)} (l_a^*)}
{ \int_1^{+\infty} dl \psi^{(-F_0)} (l)}
\label{piladeloc}
\end{eqnarray}
with
\begin{eqnarray}
\psi^{(-F_0)} (l_a) = \int_0^{+\infty} dU 
\frac{ U }{ 2 {\sqrt \pi} (\sigma l_a)^{3/2} }
e^{ - \frac{ (U + F_0 l_a)^2}{4 \sigma l_a} }
=  \frac{ F_0^2 (\sigma l_a)^{1/2}  }{ \sigma^2 2 {\sqrt \pi}  }
\int_0^{+\infty} dz z 
e^{ - l_a \frac{ F_0^2}{4 \sigma } (1+z)^2  }
\label{psiladeloc}
\end{eqnarray}
In particular, it decays exponentially as
\begin{eqnarray}
\overline{ Q_L(l_a) } = \pi_L^{(T>T_c)}(l_a) 
\oppropto_{l_a \to +\infty}  \ e^{- \frac{F_0^2}{4 \sigma} l_a } 
\label{psiladelocxi}
\end{eqnarray}
with the correlation length
\begin{eqnarray}
\xi_a = \frac{4 \sigma}{F_0^2} \oppropto_{T \to T_c^+} \frac{1}{(T-T_c)^2}
\label{xilaetoite}
\end{eqnarray}
This should be compared with the typical decay of Eq. \ref{qatyp}.
As a consequence, the finite-size scaling properties in the critical
region involves the correlation length exponent
\begin{eqnarray}
\nu_{FS}=2
\label{nuFS}
\end{eqnarray}
and not the typical correlation exponent that appears in Eq. \ref{nutydeloc}.

\subsubsection{ Sample-to-sample fluctuations in the localized phase }

In the localized phase, 
where $F_0>0$, it is the detached length $l_d^*=L-l_a^*$
that remains finite as $L \to +\infty$,
and the disorder-averaged probability of the detached length $l_d$
 can be obtained as
\begin{eqnarray}
 \overline{ P_L(l_d) } = \pi_L^{(T<T_c)}(l_a^*=L-l_d^*)
 \simeq_{L \to \infty}   \frac{\psi^{(F_0)} (l_d)}
{ \int_1^{+\infty} dl \psi^{(F_0)} (l_d)}
\label{pilaloc}
\end{eqnarray}
with
\begin{eqnarray}
\psi^{(F_0)} (l_d) = \int_0^{+\infty} dU 
\frac{ U }{ 2 {\sqrt \pi} (\sigma l_d)^{3/2} }
e^{ - \frac{ (U + F_0 l_d)^2}{4 \sigma l_d} }
\label{psilddeloc}
\end{eqnarray}
In particular, it decays exponentially as
\begin{eqnarray}
\overline{ P_L(l_d) }
 \oppropto_{l_d \to +\infty}  \ e^{- \frac{F_0^2}{4 \sigma} l_d } 
\label{pldav}
\end{eqnarray}
with the correlation length
\begin{eqnarray}
\xi_d = \frac{4 \sigma}{F_0^2} \oppropto_{T \to T_c^-} \frac{1}{(T_c-T)^2}
\label{xildetoite}
\end{eqnarray}
This should be compared with the typical decay of Eq. \ref{pdetachedtyp}.
Again, this means that the finite-size scaling properties
in the critical region are 
 governed by the correlation length exponent $\nu_{FS}=2$ of Eq. \ref{nuFS}.

\section{ Discussion }

\subsection{ Lack of self-averaging of the order parameter at criticality }

Outside criticality, the contact density is self-averaging in 
the thermodynamic limit $L \to +\infty$
\begin{eqnarray}
{\cal P}_{T<T_c}(\theta) && = \delta(\theta-1) \\
{\cal P}_{T>T_c}(\theta) && = \delta(\theta)
\label{contactnottc}
\end{eqnarray}
but exactly at criticality, we have obtained that 
the contact density $\theta$
remains distributed at criticality over the samples
 with the law of Eq.
\ref{contacttc}
\begin{eqnarray}
{\cal P}_{T_c}(\theta) = 
\frac{1}{ \pi \sqrt{ \theta (1- \theta)} }
\label{contacttcbis}
\end{eqnarray}
The lack of self-averaging of
densities of extensive thermodynamic observables
at random critical points whenever disorder is relevant
has been studied in \cite{domany95,AH,domany}.
The main idea is that off-criticality, self-averaging is ensured 
by the finiteness of the correlation function $\xi$, that allows to 
divide a big sample $L \gg \xi(T)$ into a large number
of nearly independent sub-samples. However at criticality where 
the correlation length diverges $\xi=+\infty$,
even big samples cannot be divided into nearly independent sub-samples,
and one obtains a lack of self-averaging. The cases considered in
\cite{domany95,AH,domany} were second-order phase transitions,
but the result of Eq. \ref{contacttc} means that the contact density 
is not self-averaging at the random first-order 
transition studied in this paper.

\subsection{Physical interpretation of two correlation length exponents}

The presence of two different correlation length exponents $\nu_{av}=2$
and $\nu_{typ}=1$ whenever the disorder is equivalent 
to a Brownian potential $U(x)$ is well known in other contexts,
in particular for random walks in random media
\cite{bou90,us_golosov}, for the quantum transverse field 
Ising chain \cite{danielrtfic,microcano} and more generally
in other models described by the same ``Infinite Disorder
Fixed Point'' (see the review \cite{review} and references therein).
The physical interpretation is the following \cite{danielrtfic} :

(i) the first length scale corresponds to the length $\xi_{typ}$
where the mean value $\overline{ [U(L)-U(0)] }=F_0 L $ is of order one, which yields
\begin{eqnarray}
\xi_{typ} \sim \frac{ 1}{F_0^{ \nu_{typ}} } \ \ \ {\rm with} \ \ \nu_{typ}=1
\label{explainnutyp}
\end{eqnarray}

(ii) the second length scale 
corresponds to the length $ \xi_{av}$
where most of the samples indeed have 
$(U(L)-U(0)) \sim F_0 L \pm \sqrt{ \sigma L} $
 of the same sign of the mean value,
i.e. the scale $\sqrt{ \sigma L}$ of the fluctuations 
should be of the same order of the mean value, which yields
\begin{eqnarray}
 \xi_{av} \sim \frac{ 1}{ F_0^{\nu}} \ \ \ {\rm with} \ \  \nu_{av}=2
\label{explainnuav}
\end{eqnarray}

We have found that the finite-size scaling properties 
for disorder-averaged values over the samples are governed by
$\nu_{FS}=\nu_{av}=2$ which actually saturates
the general bound $\nu_{FS} \geq 2/d_{dis}$ \cite{chayes},
where $d_{dis}$ is the dimensionality of the disorder
(here $d_{dis}=1$), whereas the typical exponent is smaller $\nu_{typ}=1$.
This possibility of a first order transition 
that remains first order in the presence of quenched disorder
has been already discussed in \cite{chayes} and in 
Sec. VII A of Ref. \cite{danielrtfic}. 

\section{ Conclusions and perspectives }

\label{secconclusion}

In this paper, we have studied the random wetting transition
 on the Cayley tree. We have obtained that the transition which is first order
in the pure case, remains first-order in the presence of disorder,
but that there exists two diverging length scales in the critical region :
the typical correlation length diverges with the exponent
 $\nu_{typ}=1$, whereas the averaged correlation length
diverges with the bigger exponent $\nu_{av}=2$ which governs
 the finite-size scaling properties over the samples.
We have describe the relations with the previously studied models 
that are governed by the same ``infinite disorder fixed point''.
We have given detailed results on the statistics of the free-energy,
and on the statistics of the contact density $\theta=l_a/L$
which constitutes the
order parameter in wetting transitions.
In particular, we have obtained that at criticality, 
the contact density is not self-averaging
but remains distributed
 over the samples in the thermodynamic limit, with the distribution
${\cal P}_{T_c}(\theta) = 1/(\pi \sqrt{ \theta (1-\theta)})$.

As explained at the beginning,
our physical motivation to consider such a model of random wetting
on the Cayley tree
was to better understand the similarities and differences
with two other types of models involving 
directed polymers and frozen disorder described in the points (i) and (ii) of 
the Introduction. The results given in the present paper suggest
the following conclusions :

(i) The freezing transition of the directed polymer in a random medium
 on the Cayley tree \cite{Der_Spo} is of a very different nature, 
since it is the
exponent $\nu=2$ which governs the free-energy singular part
(instead of $\nu_{typ}=1$ here). The appearance of a smaller
exponent $\nu'=1$ in some finite-size properties  
in the critical region \cite{Coo_Der,fssDPCT} 
should be then explained with another mechanism.

(ii) Let us now compare with the wetting and 
Poland-Scheraga model of DNA denaturation 
with a loop exponent $c>2$ that 
we have studied numerically in \cite{c2.15numerical,PStciL}.
We find that 
the random wetting model considered in the present paper
( that corresponds to the limit of loop exponent $c \to \infty$ 
 of the model studied in Ref. \cite{c2.15numerical} ) presents very
similar critical behavior : in both cases, the transition remains first order
with $\nu_{typ}=1$ and a finite contact density, but the finite-size scaling
properties over the samples involve the exponent $\nu_{FS}=2$.
We thus hope that the exact results obtained here
for the special case $c \to \infty$ will help
to better characterize the transition in 
the region of finite-loop exponent $2<c<+\infty$.

\appendix 

\section{ Properties of pure wetting transition $u_i = u_0$ }

In this Appendix, we describe the critical properties of the pure wetting transition on the Cayley tree, to compare with the disordered case considered in the text.

\subsection{ Finite-size partition function }

When all the energies $u_i$ take the same value $u_0$, 
the ratio $R_L$ of Eq. \ref{defRL} simply reads
\begin{eqnarray}
R_L^{pure} && = \left( \frac{e^{\beta u_{0} }}{K} \right)^L + \frac{(K-1) }{K} \left[ 1+ \left( \frac{e^{\beta u_{0} }}{K} \right)
+ \left( \frac{e^{\beta u_{0} }}{K} \right)^2+ \dots + \left( \frac{e^{\beta u_{0} }}{K} \right)^{L-1} \right] \nonumber \\
&& =  \left( \frac{e^{\beta u_{0} }}{K} \right)^L + \frac{(K-1) }{K}  \times \frac{ \left( \frac{e^{\beta u_{0} }}{K} \right)^L-1}{ \left( \frac{e^{\beta u_{0} }}{K} -1\right)}
\label{RLpure}
\end{eqnarray}
The critical temperature $T_c$ then corresponds 
to the point where the energy per link $u_0$ exactly compensates
the entropy per link $(\ln K)$ of the free walk
\begin{eqnarray}
e^{\beta_c u_{0} }=K \ \ \to \ \ T_c^{pure} = \frac{u_0}{\ln K}
\label{tcpure}
\end{eqnarray}

\subsection{ Delocalized phase $e^{\beta u_{0} }<K $ }

In the delocalized phase, the leading term of Eq. \ref{RLpure} remains finite
\begin{eqnarray}
R_L^{pure} (T > T_c)   \opsimeq_{L \to \infty} \frac{(K-1) }{(K - e^{\beta u_{0} } )} +...
\label{RLpuredeloc}
\end{eqnarray}
and the excess free-energy per monomer due to the wall in the thermodynamic limit $L \to +\infty$ (Eq. \ref{freediffintensif})
vanishes
\begin{eqnarray}
f_{\infty} (T>T_c) = 0
\label{freediffpuredeloc}
\end{eqnarray}
The finite-size behavior of the full free-energy difference of Eq. 
\ref{freediff}  reads near criticality 
\begin{eqnarray}
F_L^{diff}(T>T_c) = F_L (T>T_c) - F_L^{deloc} = -T \ln R_L = - T \ln \frac{(K-1) }{(K - e^{\beta u_{0} } )} \opsimeq_{ T \to T_c^+}
- T_c \ln \frac{1}{T - T_c} 
\label{freepuredeloc}
\end{eqnarray}

In the delocalized phase, the probability distribution $Q_L(l_a)$ 
of the attached length $l_a$ of Eq. \ref{Qla}
remains finite as $L \to \infty$
\begin{eqnarray}
Q_L(l_a) \equiv P_L(l_d=L-l_a ) \opsimeq_{L \to \infty}\left( 1 - \frac{e^{\beta u_{0} }}{K}  \right) 
\left( K e^{  - \beta  u_0} \right)^{-l_a} = Q_{\infty} (l_a)
\label{pdetacheddeloc}
\end{eqnarray}
It decays exponentially 
\begin{eqnarray}
Q_{\infty} (l_a)  \opsimeq_{l_a \to +\infty} e^{ - \frac{l_a}{\xi_{deloc}(T)} }
\label{qapuredelocdecay}
\end{eqnarray}
where the correlation length reads
\begin{eqnarray}
\xi^{pure} _{deloc} (T>T_c) = \frac{ 1}{u_0 ( \beta_c - \beta)} \oppropto_{ T \to T_c^+}  \frac{1}{(T-T_c)^{\nu_{pure}} }
 \ \ { \rm with \ \ } \nu_{pure}=1
\label{xipuredeloc}
\end{eqnarray}

The thermally averaged
contact density $<\theta_L>$ of Eq. \ref{thermalavthetal} vanishes as
\begin{eqnarray}
<\theta_L> \opsimeq_{L \to \infty}   \frac{1}{L} \sum_{l_a=0}^{\infty} l_a Q_{\infty}(l_a)  = \frac{1}{L (K e^{  - \beta  u_0}-1 ) }
\label{thetalpuredeloc}
\end{eqnarray}
i.e. in the critical region, it behaves as
\begin{eqnarray}
<\theta_L> \opsimeq_{L \to \infty}   = \frac{1}{L (\beta_c-\beta) }
\label{thetalpuredeloccriti}
\end{eqnarray}

\subsection{ Localized phase $e^{\beta u_{0} }>K $ }

In the localized phase $T<T_c^{pure}$,
the leading term of the partition function of Eq. \ref{RLpure}
is exponentially large in $L$
\begin{eqnarray}
R_L^{pure} (T < T_c)  \opsimeq_{L \to \infty} 
 \left( \frac{e^{\beta u_{0} }}{K} \right)^L \left[ 1+ \frac{(K-1) }
{ \left(e^{\beta u_0 } - K \right)} \right]+O(1)
\label{RLpureloc}
\end{eqnarray}
The excess free-energy per monomer due to the wall in the thermodynamic
 limit $L \to +\infty$ (Eq. \ref{freediffintensif})
reads
\begin{eqnarray}
f_{\infty} (T<T_c) =  -  ( u_0 -  T \ln K) = - (T_c^{pure} -T ) \ln K
\label{freediffpure}
\end{eqnarray}
i.e. it vanishes linearly as the temperature $T$ approaches the critical value $T_c^{pure}$.
The transition is thus first order.

The internal energy of Eq. \ref{enerdef} becomes from Eq. \ref{RLpureloc}
\begin{eqnarray}
E_L (T<T_c) \opsimeq_{L \to +\infty} - L u_0 + \frac{(K-1) u_0 e^{\beta u_0 }}
{\left(e^{\beta u_0 } - K \right) \left(e^{\beta u_0 } - 1 \right)} +...
\label{enerlocpure}
\end{eqnarray}
The intensive energy of Eq. \ref{esintensif} remains constant
 in the whole low-temperature phase
\begin{eqnarray}
e_{\infty} (T<T_c) =  -   u_0
\label{esintensifloc}
\end{eqnarray}
and presents a jump at criticality.

The detached length $l_d$ remains a finite random variable 
in the thermodynamic limit $L \to \infty$
with the following distribution ( Eq. \ref{pdetached})
\begin{eqnarray}
P_{\infty} (l_d) =  
\frac{ \delta_{l_d=0} + \theta(l_d \geq 1) 
\frac{K-1}{K} \left( \frac{e^{\beta u_{0} }}{K} \right)^{-l_d} }{ \left[ 1+ \frac{(K-1) }{K \left( \frac{e^{\beta u_{0} }}{K} -1\right)} \right]}
\label{pdetachedpureloc}
\end{eqnarray}
It decays exponentially 
\begin{eqnarray}
P_{\infty} (l_d)  \opsimeq_{l_d \to +\infty} e^{ - \frac{l_d}{\xi(T)} }
\label{pdetachedpurelocdecay}
\end{eqnarray}
where the correlation length diverges as
\begin{eqnarray}
\xi^{pure} (T<T_c) = \frac{ 1}{u_0 ( \beta - \beta_c)} \oppropto_{ T \to T_c^-}  \frac{1}{(T_c-T)^{\nu_{pure}} }
 \ \ { \rm with \ \ } \nu_{pure}=1
\label{xipureloc}
\end{eqnarray}

\subsection{ Critical point $e^{\beta_c u_{0} }=K $ }

Exactly at criticality, the ratio of Eq. \ref{RLpure} reads
\begin{eqnarray}
R_L^{pure} (T_c)  = 1 + \frac{(K-1) }{K} L
\label{RLpurecriti}
\end{eqnarray}
so that the free-energy difference is logarithmic in $L$
\begin{eqnarray}
F_L^{diff} (T_c) = F_L^{pure} (T_c) -  F_L^{deloc} (T_c)= - T_c \ln R_L = -T_c \ln \left[ 1 + \frac{(K-1) }{K} L  \right] \opsimeq_{ L \to \infty} 
-T_c \ln  L
\label{freediffcritipure}
\end{eqnarray}

The recurrence of Eq. \ref{recener} for the energy becomes
\begin{eqnarray}
E_{L+1} = \frac{ K R_L }{ K R_L +(K-1) }  (- u_{0}+E_L) =  \frac{ K +(K-1) L }{ K  +(K-1) (L+1) }  (u_{0}+E_L)
\label{recenerpurecriti}
\end{eqnarray}
leading to
\begin{eqnarray}
E_{L} ^{pure} (T_c)= -u_0 \frac{K L +(K-1) \frac{L(L-1) }{2} }{K +(K-1) L} 
=- u_0 \frac{ (L-1) }{ 2 }   \times \frac{1+ \frac{2 K}{(K-1) (L-1) } }{1+ \frac{ K}{(K-1) L}}
\label{enerpurecriti}
\end{eqnarray}
so that the energy per monomer in the thermodynamic limit $L \to \infty$ is 
\begin{eqnarray}
e_{\infty} ^{pure} (T_c)=  - \frac{ u_0 }{ 2 }   
\label{enerpurecritiintensive}
\end{eqnarray}
i.e. exactly 
in the middle of the jump between $e_{\infty} ^{pure} (T<T_c)=-u_0$ (Eq. \ref{esintensifloc}) and $e_{\infty}^{pure} (T>T_c)=0$.

The probability distribution $Q_L(l_a)$ of the attached length $l_a$
for a system of size $L$ becomes at criticality 
\begin{eqnarray}
Q_L(l_a) = \frac{ 
 \left[ \delta_{l_a,L} + \frac{(K-1)}{K} \theta(l_a < L) \right] 
}{ 1 + \frac{(K-1) }{K} L } 
\label{pdetachedpurecriti}
\end{eqnarray}
i.e. it is essentially flat over all values $0 \leq l_a \leq L$.
The two phases coexist with proportions $(l_a,l_d=L-l_a)$.
So the value of Eq. \ref{enerpurecriti} simply
 means that the domain wall between the two phases
is on average at the middle $(l_a=l_d =L/2)$, 
but the domain wall is actually anywhere on the interval 
$0 \leq l_a \leq L$ with a flat distribution.

\section{ Disordered case : transition temperatures $T_n$
 for the averaged moments $\overline{ Z_L^n}$ of the partition function }

Since in other disordered models, one sometimes 
consider the series $T_n$ of critical temperatures
for the moments $\overline{ Z_L^n}$ of the partition function, 
we discuss their properties in the present Appendix.
For $n=1$, one sees that the annealed partition
 function satisfies the recursion (see Eq. \ref{recRL})
\begin{eqnarray}
\overline{ R_{L+1} } =   \frac{ \overline{ e^{\beta u }}}{K} 
 \overline{R_L} + \frac{(K-1)}{K}
\label{recannealed}
\end{eqnarray}
This is equivalent to the pure case recursion 
 with the change $e^{\beta u_0} \to \overline{ e^{\beta u }}$
so that the annealed critical temperature is determined by the condition
(see Eq. \ref{tcpure})
\begin{eqnarray}
 \overline{ e^{\beta_1 u }} =K 
\label{tannealed}
\end{eqnarray}
i.e. in terms of the exponent $\mu(T)$ introduced in Eq. \ref{eqmu}, 
this corresponds to the condition
\begin{eqnarray}
\mu(T_1)=1
\label{mut1}
\end{eqnarray}
More generally from the power-law of Eq. \ref{powermu}, 
it is clear that the transition temperature $T_n$
for the moments $\overline{Z_l^n}$ of order $n$ 
is determined by the condition
\begin{eqnarray}
\mu(T_n)=n
\label{mutn}
\end{eqnarray}
For instance, for the case of the Gaussian distribution of Eq. \ref{gauss}, 
the expression of Eq. \ref{resmugauss}
yields 
\begin{eqnarray}
T_n=T_c \frac{1 + \sqrt{1+ 2 n \frac{\Delta^2}{T_c^2 \ln K}  }}{2}
\label{tngauss}
\end{eqnarray}


\begin{thebibliography}{99}


\bibitem{Der_Hil} B. Derrida and H. Hilhorst, J. Phys. C14, L539 (1981).

\bibitem{Cri} A. Crisanti, G. De Simone, G. Paladin and A. Vulpiani,
Physica A192, 589 (1993).

\bibitem{luck}
J.M. Luck, ``Syst\`emes d\'esordonn\'es unidimensionnels", Al\'ea
Saclay (1992) and references therein.


\bibitem{danielrtfic}
D. S. Fisher 
Phys. Rev. Lett. 69, 534-537 (1992);
D. S. Fisher 
Phys. Rev. B 51, 6411-6461 (1995).




\bibitem{Lud} 
A.W.W. Ludwig, Nucl.Phys. B330, 639 (1990)


\bibitem{review}
F. Igloi and C. Monthus, Phys. Rep. 412, 277 (2005).

\bibitem{Der_Spo}
B. Derrida and H. Spohn, J. Stat. Phys., {\bf 51}, 817 (1988).

\bibitem{Coo_Der}
J. Cook and B. Derrida, J. Stat. Phys. {\bf 63}, 505 (1991).


\bibitem{fssDPCT}
C. Monthus and T. Garel,
Phys. Rev. E 75, 051119 (2005).

\bibitem{mfisher}
M. E. Fisher, J. Stat. Phys. 34, 667 (1984).

\bibitem{Pol_Scher} D. Poland and H.A. Scheraga eds., Academic Press, New York (1970) \\
``Theory of Helix-Coil transition in Biopolymers''.

\bibitem{c2.15numerical}
T. Garel and C. Monthus, J. Stat. Mech. P06004 (2005).

\bibitem{PStciL}
C. Monthus and T. Garel, Eur. Phys. J. B 48, 393 (2005).




\bibitem{kes75}
H. Kesten, M. Koslov and F. Spitzer, Composio. Math. {\bf 30} (1975) 145.


\bibitem{sol75}
F. Solomon, Ann. Prob. {\bf 3} (1975) 1.

\bibitem{sinai}
Y. Sinai, Theor. Prob. Appl. {\bf 27} (1982) 256.

\bibitem{derrida}
B. Derrida and Y. Pomeau, Phys. Rev. Lett {\bf 48}, 627 (1982) ;
B. Derrida, J. Stat. Phys. {\bf 31}, 433 (1983).

\bibitem{afa90}
V. Afanasev, Theor. Prob. Appl. {\bf 35}, 205 (1990). 



\bibitem{bou90}
J. P. Bouchaud, A. Comtet, A. Georges and P. Le Doussal,
Ann. Phys. {\bf 201},  285(1990). 

\bibitem{bur92}
 S. F. Burlatsky, G. S. Oshanin, A. V. Mogutov, and M. Moreau
Phys. Rev. A {\bf 45 }, R6955 (1992).


\bibitem{oshanin}
G. Oshanin, A. V. Mogutov and M. Moreau,
J. Stat. Phys. {\bf 73} , 379 (1993). 
       


\bibitem{flux}
C. Monthus and A. Comtet, J. Phys. I (France) {\bf 4}, 635 (1994).

\bibitem{yor}
A. Comtet, C. Monthus and M. Yor, J. Appl. Prob. {\bf 35}, 255 (1998).


\bibitem{us_sinai}
D.S. Fisher, P. Le Doussal and C. Monthus, Phys. Rev. Lett.
  {\bf 80} 3539 (1998);
 P. Le Doussal, C. Monthus, D.S. Fisher,  Phys. Rev. E {\bf 59} 4795
(1999).

\bibitem{us_golosov}
C. Monthus and P. Le Doussal,  Phys. Rev. E {\bf 65}, 066129 (2002).

\bibitem{hilhorst}
B. Derrida and H.J. Hilhorst, J. Phys. A {\bf 16}, 7183 (1983).

\bibitem{calan}
C. Calan, J.M. Luck, T. Nieuwenhuizen and D. Petritis,
J. Phys. A {\bf 18} (1985) 501.

\bibitem{igloi98}
F. Igloi and H. Rieger.
       Phys. Rev. B {\bf 57}, 11404 (1998).



\bibitem{dharyoung}
A. Dhar and A.P. Young,  Phys. Rev. B {\bf 68}, 134441 (2003).

\bibitem{microcano}
C. Monthus, Phys. Rev. B 69, 054431 (2004).


\bibitem{domany95}
S. Wiseman and E. Domany,
Phys Rev E {\bf 52}, 3469 (1995).

\bibitem{AH}
A. Aharony, A.B. Harris,
Phys Rev Lett {\bf 77}, 3700 (1996).

\bibitem{domany}
S. Wiseman and E. Domany, 
Phys. Rev. Lett. {\bf 81} 22 (1998) and 
Phys Rev E {\bf 58} 2938 (1998).



\bibitem{chayes}
 J. T. Chayes, L. Chayes, D.S. Fisher and T. Spencer
 Phys. Rev. Lett. {\bf 57}, 2999 (1986).



\end{thebibliography}
\end{document}